\documentclass[12pt]{article}
\usepackage{graphicx}
\usepackage{comment}
\usepackage{caption}
\usepackage{keyval}
\usepackage{everysel}
\usepackage{ragged2e}
\usepackage{amssymb}
\usepackage{amsmath}
\usepackage{cite}
\usepackage{amscd}
\usepackage{subfig}
\usepackage{breqn}

\numberwithin{equation}{section}

\begin{document}
\begin{titlepage}

{\hbox to\hsize{\hfill May 2013}}

\bigskip \vspace{3\baselineskip}

\begin{center}
{\bf \large 
Neutrino Masses and Higgs Vacuum Stability} 

\bigskip

\bigskip

{\bf Archil Kobakhidze and Alexander Spencer-Smith\\ }

\smallskip

{ \small \it
ARC Centre of Excellence for Particle Physics at the Terascale, \\
School of Physics, The University of Sydney, NSW 2006, Australia \\
E-mails: archil.kobakhidze@coepp.org.au, alexss@physics.usyd.edu.au
\\}

\bigskip
 
\bigskip

\bigskip

{\large \bf Abstract}

\end{center}
\noindent 
The Standard Model electroweak vacuum has been found to be metastable, with the true stable vacuum given by a large, phenomenologically unacceptable vacuum expectation value $\approx M_{P}$. Moreover, it may be unstable in an inflationary universe. Motivated by the necessity of physics beyond the Standard Model and to accommodate non-zero neutrino masses, we investigate vacuum stability within type-II seesaw and left-right symmetric models. Our analysis is performed by solving the renormalisation group equations, carefully taking into account the relevant threshold corrections. We demonstrate that a phenomenologically viable left-right symmetric model can be constructed by matching it with the SM at one-loop. In both models we demonstrate the existence of a large area of parameter space where the Higgs vacuum is absolutely stable.

 \end{titlepage}

\section{Introduction}

The discovery of a Higgs particle with mass $m_h\approx 125-126$ GeV \cite{Aad:2012tfa,Chatrchyan:2012ufa} has sparked a resurgence of interest into the stability of the electroweak/Higgs vacuum. Under the assumption that the new particle is the Standard Model (SM) Higgs boson, the most accurate  analysis of electroweak vacuum stability in flat spacetime was performed in \cite{Degrassi:2012ry}. Stability of the Higgs vacuum appears to have strong dependence on the top quark and Higgs boson masses ($m_t$ and $m_h$ respectively) with the electroweak symmetry breaking vacuum found to be in a metastable region of $m_t -m_h$ parameter space at a confidence level of 98\%. 

With this conclusion in mind, it is worth understanding exactly what is meant by 'metastability'. Given a scalar field theory with potential bounded from below, one finds the relativistically invariant vacuum state by first minimising the classical potential, with the field vacuum expectation value (VEV)  corresponding to the minimum of the potential. Then one needs to confirm that quantum corrections to the classical potential do not significantly alter the vacuum state obtained at the classical level. Upon quantisation, all coupling constants ${\lambda_i}$ become scale-dependent effective running constants that satisfy the renormalisation group equations (RGEs). By associating renormalisation scale $\mu$ with the field value: $\mu\approx |\phi|$, the couplings are considered functions of $\phi$: $\lambda_i = \lambda_i(\phi)$. If, over the entire range of the running, the global minimum is the same as that of the classical potential, then the vacuum state is absolutely stable. For example, the electroweak vacuum is absolutely stable in the SM if the Higgs quartic coupling, $\lambda$, is positive up to the Planck scale. On the other hand, if the Higgs quartic coupling runs negative at some $\mu_I$ then the scalar potential will develop other deeper minima, with the point $\mu_I$ at which this happens the instability scale. 

If such an instability occurs, then even if, at some early time, $\phi$ attains a field value corresponding to the local minimum of the potential, it will inevitably decay into the deeper global minimum at some later time. If this decay time is greater than the age of the universe the vacuum state is metastable, if less than, then the vacuum is unstable. Thus, the concept of metastability is cosmological, and with this in mind, an analysis of the metastability bounds during inflation were performed in \cite{Kobakhidze:2013} (see also \cite{Espinosa:2007qp} for an earlier work with an alternative decay mechanism). Unless inflation takes place at sufficiently low scales, decay rates from false to true vacua were found to increase during inflation, with the region of metastable parameter space in \cite{Degrassi:2012ry} replaced by a region of unstable parameter space - the electroweak vacuum is either absolutely stable or not at all, and the only acceptable condition on the potential is that of 'absolute stability'. This may indicate the existence of new physics beyond the SM in an energy range from that currently probed at the LHC up to the instability scale $\mu_I$. 

The presence of new physics modifies the Higgs potential at higher energies by changing the $\beta$-functions for the couplings, since we now include new particles running in loops when calculating radiative corrections. New bosonic particles provide a positive contribution to the running of the Higgs quartic coupling, whist new fermionic particles contribute negatively. Examples of this effect, as applied to the Higgs potential, were given in \cite{Lebedev:2012zw,EliasMiro:2012ay} where the presence of an additional scalar singlet, introduced below $\mu_I$, was found to preserve positivity of the quartic coupling up to the Planck scale. 

Whilst one can construct many different ad hoc new physics models to resolve the vacuum stability problem \cite{Kannike:2012pe, Anchordoqui:2012fq, Allison:2012qn, Khan:2012zw, Chao:2012xt, Goudelis:2013uca, He:2013tla}, we believe that models addressing other SM  problems deserve primary consideration. In fact, empirically, we have firmly established evidence for physics beyond the SM: neutrino oscillations (and hence neutrino masses), dark matter and matter-antimatter asymmetry. Additional theoretical considerations based on the naturalness principle, such as the strong CP problem and the gauge hierarchy problem, also provide a hint of new physics beyond the SM. Amongst this evidence, neutrino mass is perhaps the most compelling, from a purely phenomenological point of view. Therefore, in this paper we analyse the effect of some models of neutrino mass generation upon stability of the Higgs vacuum.               

The rest of the paper is organised as follows. In the next section we briefly discuss the running of the Higgs quartic coupling in the SM and possible new physics which may affect this running at high energies. In Sec. 3 and 4 we present a thorough study of vacuum stability in the type-II seesaw and left-right symmetric models. Finally, in Sec. 5 the reader can find our conclusions. In Appendix A we provide details of the matching between $\overline{\text{MS}}$ and pole masses for the Higgs boson and top quark and values of the gauge couplings at the electroweak scale, while the relevant RGEs and $\beta$-functions are collected in Appendices B, C and D.

\section{Higgs vacuum stability and physics beyond the Standard Model}

As we discussed above, the Higgs vacuum stability problem may be an indication of a new physics beyond the SM. In order to discuss the kind of new physics which might be responsible for stabilisation of the Higgs vacuum, let us consider the one-loop $\beta$-function for the Higgs self-interaction coupling within the SM:
\begin{equation}
\beta _{\lambda }^{(1)}=24\lambda ^2-6y_t^4+\frac{3}{4}g_2^4+\frac{3}{8}\left(g_1^2+g_2^2\right)^2+
\lambda  \left(-9g_2^2-3g_1^2+12y_t^2\right)
\label{A1}
\end{equation}

\begin{figure}[h!]
\centering
\subfloat[$m_h = 125$ GeV]{\label{fig:SMLambdaH125}\includegraphics[width=0.49\textwidth]{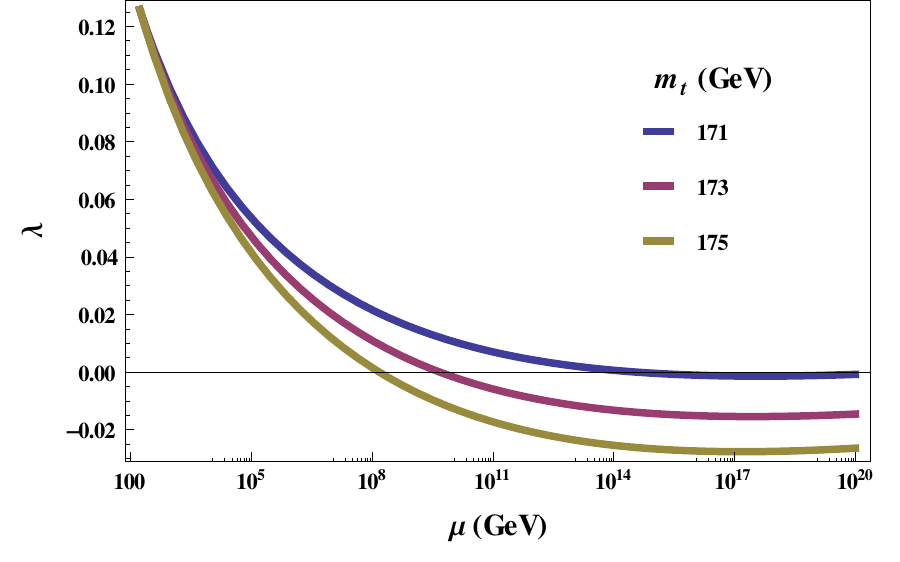}}
\subfloat[$m_h =126$ GeV]{\label{fig:SMLambdaH126}\includegraphics[width=0.49\textwidth]{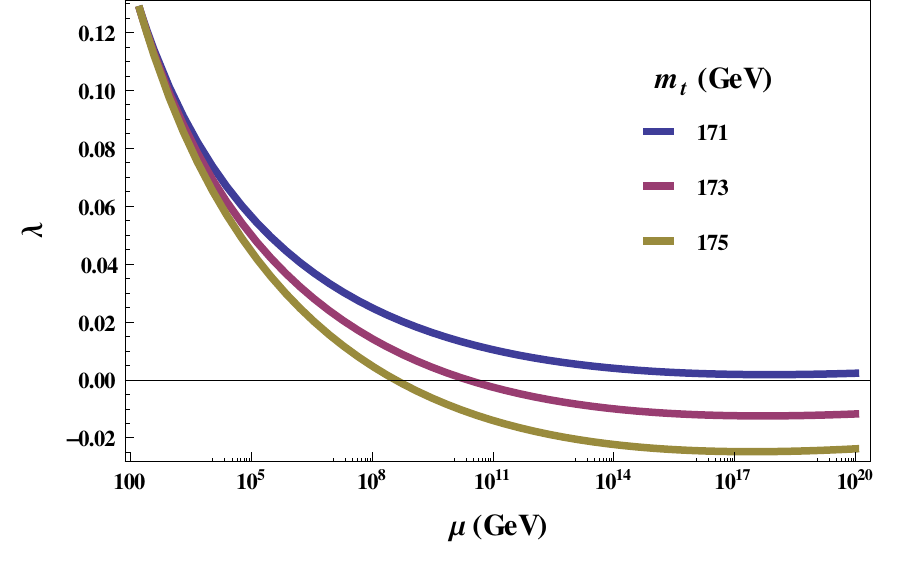}}
\caption[]{Two loop running of the Higgs quartic coupling in the SM.}
\label{fig:SMLambda}
\end{figure}

The sign of the above $\beta$-function depends on the relative strength of the self-interaction coupling $\lambda$, the top-Yukawa coupling $y_t$ and the $\text{SU}(2)_L\times \text{U}(1)_{Y}$ gauge couplings $g_2$ and $g_1$. The strength of these couplings are, in turn, defined through the Higgs boson mass, top-quark mass and masses of the weak gauge bosons, respectively, and according to the Higgs mechanism. In fact, $y_t(m_t)\approx 0.94$, $g_2(m_t)\approx 0.65$, $g_1(m_t)\approx 0.36$ and $\lambda(m_t)\approx 0.13$ so that $\beta_{\lambda}<0$ around the electroweak scale. This drives the effective running self-interaction coupling $\lambda$ to negative values at higher energies $\mu_I\approx 10^{10}$ GeV (see Figure \ref{fig:SMLambda}), rendering the electroweak vacuum unstable. Therefore, if new physics stabilises the electroweak vacuum it must enter into the game at energies below the instability scale, $\mu\lesssim \mu_I$. There are three distinct logical possibilities for such new physics: (i)~Extensions that influence the Higgs self-interaction coupling at higher energies; (ii)~ Extensions that influence the top-Yukawa interaction coupling at higher energies; and (iii)~Embeddings of the $SU(2)_L\times U(1)_Y$ electroweak gauge group into a wider group $G$ that influences the electroweak gauge couplings at higher energies. 

Working within mass-independent subtraction schemes, such as the $\overline{\text{MS}}$  scheme, the behaviour of an effective coupling is modified by new physics in two ways. The first is a finite threshold correction that matches effective couplings at low and higher energies at a characteristic matching scale where the new physics kicks in. The second is a modification of the corresponding $\beta$-function by the new particles and interactions. Models with an extended Higgs sector are shown to be capable of resolving the vacuum stability problem due to the heavy scalar threshold effect \cite{Lebedev:2012zw, EliasMiro:2012ay} or due to the extra positive contribution to $\beta_{\lambda}$ (\ref{A1}). Extra electroweak fermions may drive the (weak at low energies) $g_1$ and $g_2$ couplings to be strong enough to reverse the sign of $\beta_{\lambda}$ (\ref{A1}) from negative to positive at high energies. Other variations of these models have been discussed in \cite{Kannike:2012pe, Anchordoqui:2012fq, Allison:2012qn, Khan:2012zw, Chao:2012xt, Goudelis:2013uca, He:2013tla}. However, to the best of our knowledge, other interesting options such as models with fermionic and/or gauge threshold effects have not been considered so far, and we will present such models below.

The observation of massive neutrinos through their flavour oscillations provides our prime motivation for new physics beyond the SM. The simplest mechanism for neutrino mass generation is the type-I seesaw model \cite{Minkowski:1977sc, Yanagida:1979wu, Gell-Mann:1979sg, Mohapatra:1979ia} which introduces extra electroweak singlets of massive right-handed neutrinos. These neutrinos interact with ordinary neutrinos through the Higgs field, in a manner similar to that of the Higgs-top Yukawa interaction. Consequently, the only effect of these new particles upon vacuum stability comes from top Yukawa-type contributions to $\beta_{\lambda}$. Therefore, the vacuum stability problem can not be resolved within the simplest type-I seesaw models, as demonstrated in \cite{Casas:1999cd, Rodejohann:2012px, Chakrabortty:2012np}. 

The type-III seesaw model \cite{Foot:1988aq} employs an electroweak triplet of fermions instead of the fermionic singlet of the type-I seesaw model. These fermions modify the running of the effective gauge coupling $g_2$ such that $\beta_{\lambda}$ (\ref{A1}) receives a positive contribution and changes its sign (to positive). For this to happen the triplet must be sufficiently light \cite{Gogoladze:2008ak, Chen:2012faa, He:2012ub}, which undermines the prime motivation for the seesaw mechanism.

Finally, the type-II seesaw mechanism \cite{Schechter:1980gr, Magg:1980ut, Cheng:1980qt, Lazarides:1980nt, Mohapatra:1980yp} introduces an electroweak triplet of scalars, which can be quite heavy. Here we have two effects that potentially solve the Higgs vacuum stability problem. One is a positive contribution to $\beta_{\lambda}$ from Higgs-scalar triplet interactions and the other is a finite threshold effect. All the previous studies \cite{Gogoladze:2008gf, Arina:2012fb, Chun:2012jw, Chao:2012mx}, with the exception of \cite{Dev:2013ff}, considered only the first effect, which is significant for low scalar triplet masses. For heavy triplet scalar mass the finite threshold effect becomes more important and it cannot be ignored. The authors of Ref. \cite{Dev:2013ff} discuss the threshold correction, but do not implement it correctly and try to make it small under the false impression that the correction reduces the instability scale and are forced to explore a region of parameter space for which the couplings are non-perturbative. In our work we correctly implement the threshold correction at all scales, and in particular for large values of $m_\Delta$, where it is non-negligible.

The seesaw mechanism can be extended to explain the mass hierarchies of quarks and charged leptons \cite{Berezhiani:1983hm, Berezhiani:1985in, Dimopoulos:1983rz}. In these  universal seesaw models, all the SM Yukawa couplings are just a low-energy manifestation of more "fundamental" Yukawa interactions involving additional fermions and bosons. Thus, above the relevant mass scale, the SM Yukawa couplings get modified by finite threshold effects and renormalisation group running.  Particularly interesting are the left-right symmetric models with extended electroweak gauge group $\text{SU}(2)_L\times \text{SU}(2)_R\times \text{U}(1)_{B-L}$ \cite{Davidson:1987mh, Davidson:1987tr, Chang:1986bp, Rajpoot:1986nv, Davidson:1989bx, Berezhiani:1991ds}. Here, the SM hypercharge gauge coupling $g_1$ and top-Yukawa coupling $y_t$ also receive  threshold corrections once the SM is properly matched with the left-right symmetric model at high energies. We will demonstrate this within the 'alternative' left-right symmetric model (ALRSM) with a minimal scalar sector comprising of only $SU(2)_L$ and $SU(2)_R$ doublets of scalar fields. In previous studies either explicit left-right symmetry breaking or further extension of the scalar sector have been employed to overcome some phenomenological difficulties. We will show that the ALRSM is phenomenologically viable on its own and provides a framework for solving the vacuum stability problem. 

\section{Type-II Seesaw Model}

\subsection{The Model}

In this section we would like to investigate vacuum stability in an extension of the SM which incorporates a minimal type-II seesaw mechanism for neutrino mass generation. We extend the scalar sector of the SM by adding a scalar $\text{SU}(3) \times \text{SU}(2) _L\times \text{U}(1)_Y$ triplet, $\Delta \in (\boldsymbol{1},\boldsymbol{3},1)$, to the ordinary SM electroweak doublet, $\phi \in (\boldsymbol{1},\boldsymbol{2},1/2)$. The hypercharge is normalised so that $Q = T_{3L} + Y$. The most general renormalisable tree-level scalar potential is
\begin{multline}
\label{Scalar Pot}
V(\phi,\Delta) = -m_\phi^2 \phi^\dag \phi + \lambda (\phi^\dag \phi)^2 +m_\Delta^2 \text{tr}(\Delta^\dag \Delta) + \frac{\lambda_1}{2}(\text{tr}(\Delta^\dag \Delta))^2
\\ + \frac{\lambda_2}{2} \left[  (\text{tr}(\Delta^\dag \Delta))^2 - \text{tr}(\Delta^\dag \Delta)^2 \right] +\lambda_4 (\phi^\dag \phi)\text{tr}(\Delta^\dag \Delta) + \lambda_5 \phi^\dag [\Delta^\dag , \Delta]\phi + \left[ \frac{\lambda_6}{\sqrt{2}} \phi^T i\sigma_2 \Delta^\dag \phi + \text{h.c.} \right].
\end{multline}
The SM gauge group admits the following gauge invariant Yukawa interaction between the triplet and left-handed leptons
\begin{equation}
\label{Neutrino Majorana}
\frac{1}{\sqrt{2}}(y_\Delta)_{fg}l_L^{Tf}C i \sigma_2 \Delta l_L^g + h.c.
\end{equation}
Note that the normalisation for this Yukawa coupling was chosen to agree with \cite{Schmidt:2007nq}, however we chose all other SM Yukawa couplings to be unnormalised. The potential \eqref{Scalar Pot}, along with the VEV for the doublet $\langle \phi \rangle = v_{EW}/\sqrt{2}$, admits a small but non-zero VEV for the triplet, $\langle \Delta \rangle = v_{\Delta} \ll v_{EW}$. As a result, small Majorana masses are generated by the Yukawa interaction \eqref{Neutrino Majorana}
\begin{equation}
\label{Neutrino Mass}
m_\nu= v_\Delta \frac{y_\Delta}{\sqrt{2}} = \frac{v_{EW}^2 \lambda_6}{2 m_\Delta^2} y_\Delta.
\end{equation}
The presence of a scalar triplet breaks the $SU(2)$ custodial symmetry of the SM and thus modifies the $\rho$-parameter. Experimental bounds on $\rho$ provide the constraint  \cite{Beringer:1900zz}
\begin{equation}
\label{Rho Constraint}
v_\Delta \lesssim 1 \text{GeV}.
\end{equation}
.
\subsection{Matching the SM with the Type-II Seesaw Model}

As is well known, the RGEs take a particularly simple form in mass-independent subtraction schemes, such as $\overline{\text{MS}}$. However, we pay a price for this convenience - high and low energy physics do not manifestly decouple. To circumvent this problem one must carefully match the high energy theory with a low energy effective theory at a relevant matching scale. To this end, we treat the SM as the low energy effective theory of the type-II model described above, with an effective Higgs quartic coupling, $\lambda_h$, valid below the scale set by $m_\Delta$. By integrating out the heavy scalar triplet in the tree-level approximation, we obtain the following relation between $\lambda_h$ and parameters in the potential \eqref{Scalar Pot} at the scale $\mu = m_\Delta$
\begin{equation}
\label{Shift}
\lambda_{h} = \lambda - \frac{\lambda_{6}^2}{2m_\Delta^2}.
\end{equation}

\subsection{Boundary Conditions for the RGEs}

To solve the RGEs we must define the relevant coupling constants at some given scale. The SM gauge coupling constants, $g_1$, $g_2$ and $g_3$ can be taken from measurements at the Z-pole, whilst the top Yukawa coupling, $y_t$, and Higgs quartic coupling, $\lambda_h$ can be inferred from top quark and Higgs boson mass measurements (see Appendix \ref{app:Initial Conditions} for details). These five are the dominant SM couplings and we ignore the other SM couplings in our analysis. 

We evaluate these five SM couplings at the top pole-mass, $m_t$ and run the couplings from there. Above $m_\Delta$ we evolve all of the couplings according to the RGEs for the full theory, with $\lambda_h$ matched to $\lambda$ by enforcing the discontinuous shift \eqref{Shift}. Note that this threshold correction immediately \emph{increases} the instability scale, since we run $\lambda_h$ from $m_t$, ensuring it remains positive, then we switch to $\lambda$, the self-coupling for the full theory, which receives a positive correction thanks to \eqref{Shift}. For the full theory we use the one-loop $\beta^{(1)}_i$ from \cite{Schmidt:2007nq}, and the two-loop $\beta^{(2)}_i$ from the SM. 

We must also specify boundary conditions for all the couplings not appearing in the SM -  the set $\{y_\Delta, \lambda_1, \lambda_2, \lambda_4, \lambda_5, \lambda_6 \}$ and the scale, $m_\Delta$, at which we define these. Constraints on these arising from the condition of absolute stability of the type-II seesaw scalar potential will be discussed in the next section. However, at this point we should note that our choice of boundary values for the coupling constants should satisfy the stability conditions and also ensure that they remain satisfied over the entire running. In order to trust our RGEs we must ensure that perturbation theory does not break down, so for all dimensionless couplings, $\lambda_i$ we require $\lambda_i < \sqrt{4\pi}$ for both the boundary condition and over the range of the running. 

Boundary conditions for the coupling constants must also satisfy a set of phenomenological and theoretical bounds, as well as bounds set by the requirements of stability and perturbativity. The constraint on $v_\Delta$ \eqref{Rho Constraint} implies
\begin{equation}
\label{Rho Bound}
\frac{v_\Delta}{v_{EW}} = \frac{v_{EW}\lambda_6}{m_\Delta^2} < 0.01.
\end{equation}
So if the dimensionful coupling, $\lambda_6$, is some multiple of the triplet mass: $\lambda_6 = K_{m_\Delta} m_\Delta$, then for any given $m_\Delta$ we must have 
\begin{equation}
K_{m_\Delta} <  0.01 \frac{m_\Delta}{v_{EW}}.
\end{equation} 
This bound is only particularly relevant for low triplet masses, where, for example, with $m_\Delta = 1$ TeV we find $K_{1TeV} < 0.04$. For larger $m_\Delta$ the bound coming from the $\rho$-parameter relaxes and $\lambda_6$ can be quite large. In this case $\lambda$ can receive a large threshold correction (see \eqref{Shift}). Then the bound on $\lambda_6$ coming from the requirement of perturbativity of $\lambda$ becomes more stringent than \eqref{Rho Bound}.

Since we are working in a type-II seesaw scenario we would like light neutrinos! Given a choice of $m_\Delta$, and taking into account the bounds on $\lambda_6$ from the $\rho$-parameter, we are able to get light neutrinos by taking the coupling $y_\Delta$ to be very small. Given a choice of $m_\Delta$ and with $m_\nu \approx 0.1\text{eV}$ we determine $y_\Delta$ from \eqref{Neutrino Mass}, which removes $y_\Delta$ as a free parameter of the theory.

The remaining dimensionless parameters in the full type-II model are free and not constrained by low energy observables, hence we have a rather large theoretical degeneracy. Since we do not aim to study the model in full phenomenological detail here, we make the simple and natural assumption that all the dimensionless couplings in the scalar potential are of the same order of magnitude. In fact, we take
\begin{equation}
\lambda_1(m_\Delta) = \lambda_2(m_\Delta) = \lambda_4(m_\Delta) = \lambda_5(m_\Delta) = \lambda(m_\Delta).
\end{equation}
This serves as a nice way to remove some more free parameters in the theory - the only input values are now the scale $m_\Delta$, and the bound on $\lambda_6$ which arises from this choice of scale. 

From Figures \ref{fig:SMLambdaH125}, \ref{fig:SMLambdaH126} one can see that larger values of $m_t$ result in lower values of the instability scale. For the maximum reasonable top pole-mass, $m_t \approx 175$ GeV, the instability scale occurs at around $10^8 \text{GeV}$, so we make this the maximum value for $m_\Delta$. 

\subsection{Stability Conditions}

Our primary objective is to find a set of boundary conditions for the coupling constants that result in an absolutely stable Higgs potential. In the SM the condition of absolute stability is equivalent to the requirement that the Higgs quartic coupling $\lambda$ remain positive all the way up to the Planck scale. When we switch to the full type-II seesaw theory there are additional stability conditions, and also requirements for the absence of tachyonic modes. These conditions were derived in \cite{Arhrib:2011uy}, but the potential used there is of a different form to that in \cite{Schmidt:2007nq}. For the couplings in \eqref{Scalar Pot} the stability conditions are
\begin{gather}
\label{Stability Conditions}
\lambda > 0,\\
\lambda_1 > 0,\\
\lambda_1 + \frac{\lambda_2}{2} > 0,\\
\lambda_4 \pm \lambda_5 + 2\sqrt{\lambda \lambda_1} > 0,\\
\lambda_4 \pm \lambda_5 + 2\sqrt{\lambda \left(\lambda_1 + \frac{\lambda_2}{2}\right)} > 0.
\label{End Stability Conditions}
\end{gather}
The conditions for the absence of tachyonic modes are
\begin{gather}
\label{Tachyonic Modes}
\lambda_6 > 0,\\
-\lambda_5 v_\Delta < \lambda_6,\\
-2\lambda_5 v_\Delta - \frac{\lambda_2 v_\Delta^3}{v_{EW}^2} < \lambda_6,
\label{End Tachyonic Modes}
\end{gather}
Whilst our choice of boundary conditions should satisfy these, they must also hold for all values of the running, so part of the analysis includes checking that these conditions hold all the way up to the Planck scale.

\subsection{Solution to the RGEs}

\begin{figure}[h!]
\centering
\subfloat[$m_\Delta =1\text{ TeV}$, $\lambda_6 = 0.01 m_\Delta$]{\label{fig:LambdaMDel1TeVH125 Mu001MDel}\includegraphics[width=0.49\textwidth]{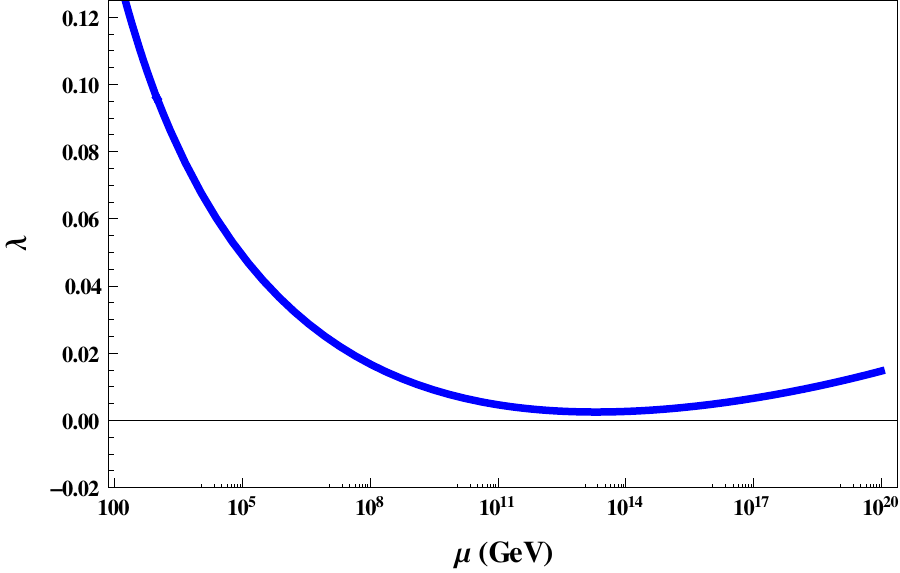}}
\subfloat[$m_\Delta =100\text{ TeV}$, $\lambda_6 = 0.13 m_\Delta$.]{\label{fig:LambdaMDel100TeVH125Mu013MDel}\includegraphics[width=0.49\textwidth]{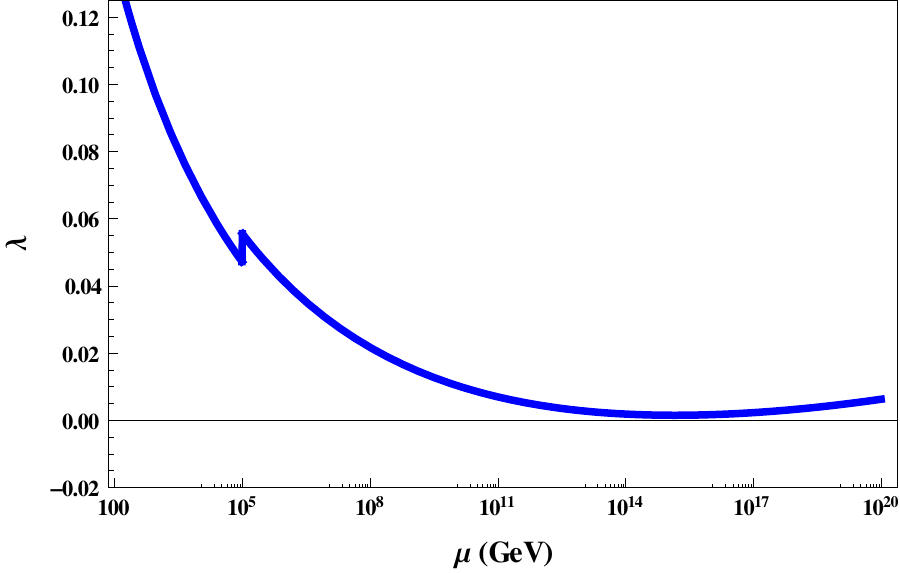}} \qquad
\subfloat[$m_\Delta =10^8\text{GeV}$, $\lambda_6 = 0.17 m_\Delta$.]{\label{fig:LambdaMDel10to8GeVH125Mu017MDel}\includegraphics[width=0.49\textwidth]{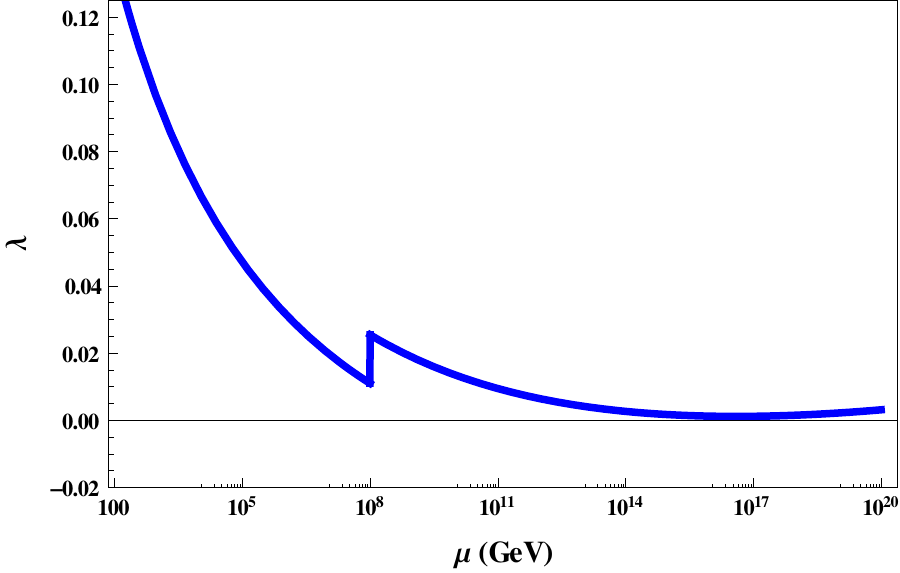}}
\subfloat[$m_\Delta =100\text{ TeV}$, $\lambda_6 = 0.57 m_\Delta$.]{\label{fig:LambdaMDel100TeVH125Mu057MDel}\includegraphics[width=0.49\textwidth]{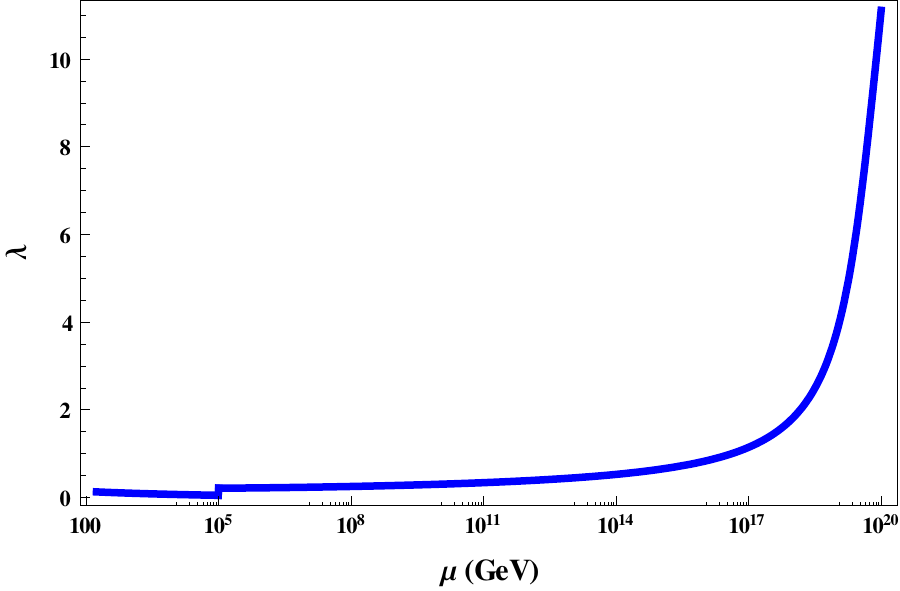}}
\caption[]{One loop running of the Higgs quartic coupling in the type-II seesaw model, with $m_h = 125$ GeV and $m_t = 173$ GeV.}
\label{fig:LambdaH125}
\end{figure}

\begin{figure}[h!]
\centering
\subfloat[$m_\Delta =10^8\text{GeV}$, $\lambda_6 = 0.17 m_\Delta$.]{\label{fig:StabCondsMDel10to8GeVH125Mu017MDel}\includegraphics[width=0.49\textwidth]{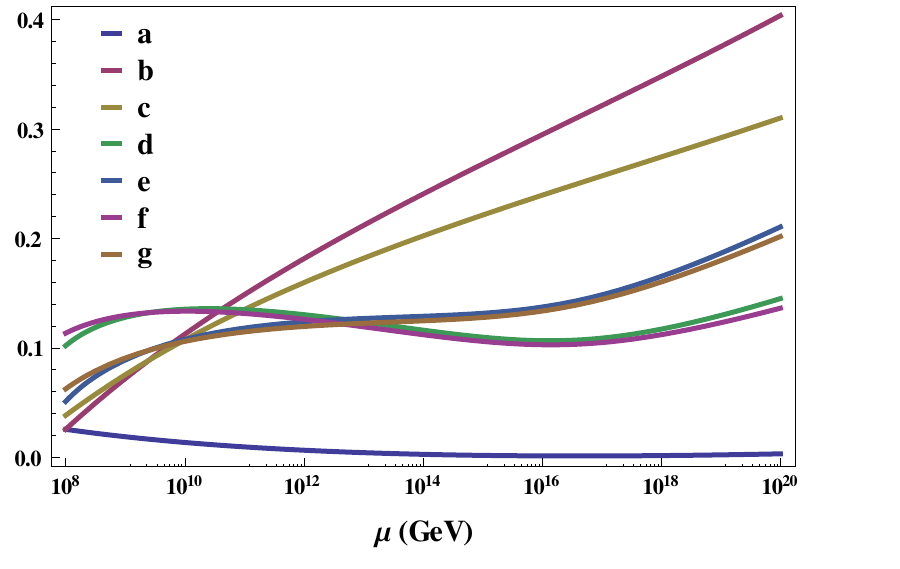}}
\subfloat[$m_\Delta =1\text{ TeV}$, $\lambda_6 = 0.01 m_\Delta$.]{\label{fig:TachCondsMDel1TeVH125Mu001MDel}\includegraphics[width=0.49\textwidth]{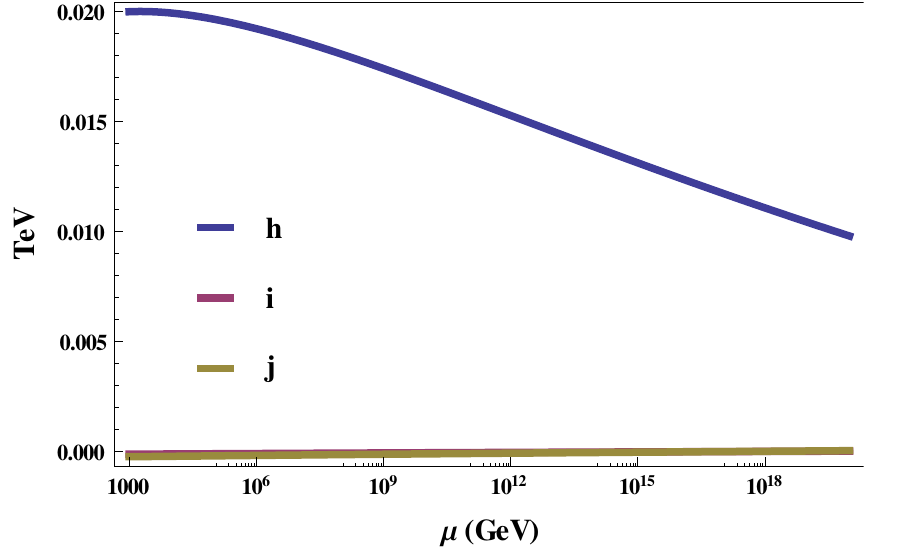}}
\caption[]{Conditions for stability and absence of tachyonic modes in the scalar potential of the type-II seesaw model, with $m_h = 125$ GeV and $m_t = 175$ GeV. Line labels correspond to the stability conditions \eqref{Stability Conditions}-\eqref{End Tachyonic Modes} with $a \equiv \lambda$, $b \equiv \lambda_1$, $c \equiv \lambda_1 + \frac{\lambda_2}{2}$, $d/e \equiv \lambda_4 \pm \lambda_5 + 2\sqrt{\lambda \lambda_1}$, $f/g \equiv \lambda_4 \pm \lambda_5 + 2\sqrt{\lambda \left(\lambda_1 + \frac{\lambda_2}{2}\right)}$, $h \equiv \lambda_6$, $i \equiv -\lambda_5 v_\Delta$, $j \equiv -2\lambda_5 v_\Delta - \frac{\lambda_2 v_\Delta^3}{v_{EW}^2}$.}
\label{fig:StabConds}
\end{figure}

Our solutions to the RGEs are presented graphically in Figures \ref{fig:LambdaH125} and \ref{fig:StabConds}. Namely, Figures \ref{fig:LambdaMDel1TeVH125 Mu001MDel} to \ref{fig:LambdaMDel10to8GeVH125Mu017MDel} show the effect of introducing the type-II seesaw scalar triplet, $\Delta$, with a range of masses from $1$ TeV up to the instability scale, $\mu_I \approx 10^8$ GeV. Values of $\lambda_6$ have been chosen at the minimum value which results in an absolutely stable Higgs potential for $m_h = 125$ GeV, $m_t = 173$ GeV and the gauge couplings at their mean world average values, as measured at the Z-pole (see appendix A). An exception is Figure \ref{fig:LambdaMDel100TeVH125Mu057MDel}, which shows how the perturbative bound of $\lambda < \sqrt{4\pi}$ is violated if one chooses too large an initial value for $\lambda_6$ (whilst still satisfying phenomenological bounds as discussed in the previous section).

As we can see, there are two major contributions that determine the running of $\lambda$ at high energies - one is the modification of the $\beta$-functions for scales $\mu > m_\Delta$ and the second is the finite shift \eqref{Shift} due to the threshold correction at $\mu = m_\Delta$. The threshold correction depends on the ratio $\lambda_6/m_\Delta$ and since $\lambda_6$ must always be positive for the absence of tachyonic modes (see \eqref{Tachyonic Modes}), we always have an improved stability bound in the model. The effect arising from modification of the $\beta$-function is more important for lighter triplet masses (TeV scale), as the modified running occurs over a greater distance whilst the bound on $\lambda_6$ from \eqref{Rho Bound} means the threshold correction can only be small. However, for larger triplet masses, the threshold correction is the most important - the range of modified running is reduced, but the threshold correction can now be quite large. For mid-scale $m_\Delta \approx 10^5$ GeV models both effects contribute, with $\lambda_6$ relatively unconstrained. For small enough $m_\Delta$ and large enough $\lambda_6$ the sign of $\beta_\lambda$ changes and $\lambda$ increases at high energies. This is demonstrated in Figure \ref{fig:LambdaMDel100TeVH125Mu057MDel} where $\lambda$ hits the Landau pole at $\mu \approx 10^{18}$ GeV.

We also verified the conditions \eqref{Stability Conditions}-\eqref{End Tachyonic Modes}. Figures \ref{fig:StabCondsMDel10to8GeVH125Mu017MDel} and \ref{fig:TachCondsMDel1TeVH125Mu001MDel} are examples of, respectively, the stability conditions and conditions for the absence of tachyonic modes, plotted over the range that the couplings run. These two particular examples ware chosen as they represent the worst cases over all the different possible boundary conditions. It's clear from the figures that these conditions are always met for any choice of initial values that also satisfy the phenomenological and theoretical bounds given above.


\section{Alternative Left-Right Symmetric Model}

\subsection{The Model}

The SM is a chiral theory in which parity is explicitly broken (for SU($2$)) by hand. Left-right symmetric models are a class of models in which parity is restored as a symmetry of the full theory, and in which parity is then spontaneously broken \cite{Mohapatra:1974gc} (for a nice review see \cite{Senjanovic:2011zz}). The gauge group of the left-right symmetric model is $\text{SU}(3) \times \text{SU}(2)_L \times \text{SU}(2)_R \times \text{U}(1)_{B-L}$, which undergoes a chain of spontaneous symmetry breaking $\text{SU}(3) \times \text{SU}(2)_L \times \text{SU}(2)_R \times \text{U}(1)_{B-L} \rightarrow \text{SU}(3) \times \text{SU}(2)_L \times \text{U}(1)_Y \rightarrow \text{SU}(3) \times  \text{U}(1)_{EM}$. Fermionic representations are also symmetric under $L \leftrightarrow R$, so, as a necessity, right hand neutrinos are included in these models. Other than providing an explanation for parity breaking in the SM, the automatic inclusion of a right handed neutrino provides the main motivation for considering left-right symmetric models, with the type-I seesaw mechanism now a natural explanation for the large mass hierarchy between the neutrinos and charged leptons. In fact, the restoration of parity and inclusion of right handed neutrinos both come into play here, since the neutrino masses are related to the scale of symmetry breaking. Adopting the U(1) normalisation of the SM, (B-L) quantum numbers for any representation are determined via the relationship
\begin{equation}
\label{normalisation}
Q = (T_3)_L + (T_3)_R + (B-L) = (T_3)_L + Y
\end{equation}
The SM quarks and leptons, with the addition of a right handed neutrino, fit into the following representations of the left-right symmetric gauge group
\begin{gather}
q_L \in (\mathbf{3},\mathbf{2},\mathbf{1},1/6),
\\ q_R \in (\mathbf{3},\mathbf{1},\mathbf{2},1/6),
\\ l_L \in (\mathbf{1},\mathbf{2},\mathbf{1},-1/2),
\\l_R \in (\mathbf{1},\mathbf{1},\mathbf{2},-1/2).
\end{gather}
Since we do not observe any gauge bosons for the right handed gauge group the left-right symmetry must be broken, with $M_{W_R} \gg M_{W_L}$. Current experimental bounds on the masses of $Z_R$ of $M_{Z_R} \gtrsim 1.2$ TeV \cite{Beringer:1900zz} dictate that we must have a large hierarchy of scales between the breaking of left-right symmetry and the electroweak scale: $v_{EW} \ll v_R$.  In order to break the gauge group down to the SM one requires a set of scalars which will break $\text{SU}(2)_R \times \text{U}(1)_{B-L} \rightarrow \text{U}(1)_Y$. This symmetry breaking pattern is identical to that of electroweak symmetry breaking in the SM ($\text{SU}(2)_L \times \text{U}(1)_{Y} \rightarrow \text{U}(1)_{EM}$.) so we break with two sets of complex scalars
\begin{gather}
\phi_L \in (\mathbf{1},\mathbf{1},\mathbf{2}, 1/2),
\\ \phi_R \in (\mathbf{1},\mathbf{2},\mathbf{1}, 1/2).
\end{gather}
The B-L charges of these doublets are unimportant at this point, but will be required later for the fermionic couplings. The scalar potential is
\begin{equation}
\label{scalar pot}
V(\phi_L,\phi_R) = -m^2 \left( \phi_L^\dag \phi_L + \phi_R^\dag \phi_R \right) + \frac{\lambda}{2} \left(\phi_L^\dag \phi_L + \phi_R^\dag \phi_R \right)^2  + \sigma \phi_L^\dag \phi_L\phi_R^\dag \phi_R.
\end{equation}
Analysis of the vacua of this potential shows that the potential is bounded from below if $\lambda > 0$ and $\sigma > -2\lambda$. In order to break L-R symmetry we simply require $\sigma > 0$, which, given the first condition, automatically satisfies the second. In this case we find $v_R^2 = \frac{m^2}{\lambda}$ and $v_L = 0$. We need some scalar with a non-zero VEV to perform the job of the Higgs in the SM - there are two ways to proceed. The first option, most usually taken, is to add to the model a scalar bidoublet that attains a non-zero VEV, breaking the remaining $SU(2)_L$ symmetry, and which also couples to the fermions. This is usually referred to as the 'minimal' left-right symmetric model. The second option is to add no further scalars to the theory and hope that $\phi_L$ attains a VEV via radiative corrections to the potential. If so, then it's possible for $\phi_L$ to play the role of the Higgs. This is referred to as the 'alternative' left-right symmetric model (ALRSM) \cite{Davidson:1987mh}. The scalars couple to the SM fermions, $f_i$, via dimension five effective operators, which reduce to the SM Yukawa couplings below the scale $v_R$ when $\phi_R$ attains it's VEV. Operators of the form
\begin{equation}
\label{Majorana}
\frac{y_{ij}^{L}}{m_N}(l_{iL}\phi_L)(l_{jL}\phi_L) + \frac{y_{ij}^R}{m_N}(l_{iR}\phi_R)(l_{jR}\phi_R) + \text{h.c.} 
\end{equation}
result in Majorana masses for the neutrinos, whist
\begin{equation}
\label{Dirac}
\frac{y^{D}_{ij}}{m_D}(\overline{f_{iL}}\phi_L^*)(f_{jR}\phi_R) + h.c., \ f=q,l.,
\end{equation}
give Dirac masses. 
The full renormalisable theory contains extra vector-like fermionic states which, when integrated out, result in the operators \eqref{Majorana} and \eqref{Dirac}. We know from the analysis of vacuum stability in type-II seesaw models that the presence of extra scalars acts to stabilise the Higgs potential, so we choose to pursue the ALRSM, since it contains the minimal left-right symmetric scalar content, and also includes extra fermions, thus representing a 'worst-case' scenario (in terms of vacuum stability) for this class of models.

To reproduce, for example, the top-Yukawa coupling at low energy, we introduce two colour triplets of fermions, $T_{i L},\ T_{i R}$ with interaction terms
\begin{equation}
\label{top coupling}
y_T \overline{q^{(3)}_{iR}}\tilde{\phi}_R T_{iL} + y_T\overline{q^{(3)}_{iL}}\tilde{\phi}_L T_{iR} + M_T \overline{T_L}T_R + h.c.
\end{equation}
Similarly, each SM fermion, $f_i$, has a vector-like partner, $F_i$, transforming as
\begin{gather}
\label{vector-like}
N_L, N_R \in (\mathbf{1},\mathbf{1},\mathbf{1}, 0), \\
E_L, E_R \in (\mathbf{1},\mathbf{1},\mathbf{1}, -1), \\
U_{iL}, U_{iR} \in (\mathbf{3},\mathbf{1},\mathbf{1}, 2/3), \\
D_{iL}, D_{iR} \in (\mathbf{3},\mathbf{1},\mathbf{1}, -1/3),
\end{gather}
with this list duplicated for each generation. These have masses 
\begin{equation}
M_{F_i}, F_i=N,E,U,D,\ldots, N_\tau, \boldsymbol{\tau}, T,B,
\end{equation} 
which, \emph{a priori}, we have no reason to expect to be anything other than arbitrary. However, again looking at neutrino masses for some kind of motivation, we implement a  universal seesaw mechanism \cite{Davidson:1987mh} with the hierarchy of vector-like fermion masses the inverse of the chiral fermions in the SM
\begin{equation}
\label{hierarchy}
\frac{m_{f_i}}{m_{f_j}} = \frac{M_{F_j}}{M_{F_i}}.
\end{equation}

\subsection{Tree Level Threshold Corrections}

As we did for our analysis of vacuum stability in the type-II seesaw model, we will use the RGEs obtained from dimensional regularisation and the $\overline{\text{MS}}$ scheme. This requires us to work in the effective field theory picture, integrating out heavy particles at their mass threshold and matching couplings as we go. The tree level threshold corrections are as follows.

\subsubsection{Gauge couplings}

We match the gauge couplings when left-right symmetry is broken i.e. at $v_R$. We chose the $\text{U}(1)_{B-L}$ normalisation to match that of the SM (see \eqref{normalisation}). Thus, the relation between $g_R$, $g_{B-L}$ and $g_1$ when $\text{SU}(2)_R \times \text{U}(1)_{B-L} \rightarrow \text{U}(1)_Y$ is exactly the same as the relation between  $g_2$, $g_1$ and $e$ when $\text{SU}(2)_L \times \text{U}(1)_{Y} \rightarrow \text{U}(1)_{EM}$ in the SM. So the matching condition for $g_{B-L}$ is
\begin{equation}
\label{U(1) matching}
g_1 = \frac{g_R \ g_{B-L}}{\sqrt{g_R^2 + g_{B-L}^2}} = \frac{g_2 \ g_{B-L}}{\sqrt{g_2^2+g_{B-L}^2}},
\end{equation}
since $g_R = g_L = g_2$ thanks to left-right symmetry. The $SU(3) \times \text{SU}(2)_L$ symmetry is unbroken over the range of the running, so there are no matching conditions for $g_3$ or $g_L = g_2$. Since we work in $\overline{\text{MS}}$, there are no corrections to the gauge couplings at a heavy fermion threshold (in fact, to all orders, not just at tree level).

\subsubsection{Scalar couplings}

The choice of normalisation for the quartic coupling in \eqref{scalar pot} differs from that in the SM by a factor of two, so at the first heavy particle threshold after the top quark, we must take
\begin{equation}
\lambda_{eff} = 2\lambda_h
\end{equation}
where $\lambda_h$ is the quartic coupling for the Higgs in the SM. 

There are no threshold corrections to $\lambda$ arising from integrating out a heavy fermion at tree-level, but we find a non-trivial matching condition from integrating out the heavy right handed scalars. To integrate out the heavy physical field we first shift the right handed scalar field by it's VEV 
\begin{gather}
\phi_L = \frac{1}{\sqrt{2}} \left(\begin{array}{c}\varphi_1 + i \varphi_2 \\ h  + i\varphi_3\end{array}\right), \ \ \phi_R = \frac{1}{\sqrt{2}} \left(\begin{array}{c}\varphi_5 + i \varphi_6 \\ H + \sqrt{2}v_R + i\varphi_7\end{array}\right),
\end{gather}
then take the unitary gauge, in which case the potential reads
\begin{multline}
\label{physical potential}
V(h,H) = \frac{1}{2} \frac{\sigma m^2}{\lambda} h^2 + m^2 H^2 + \frac{\lambda}{8}\left( h^4 + H^4 \right)\\ + \frac{\lambda + \sigma}{4}H^2 h^2 + \frac{\lambda + \sigma}{\sqrt{2\lambda}} m H h^2 + m\sqrt{\frac{\lambda}{2}}H^3.
\end{multline}
The idea is to match the low energy limit of this scalar potential with that of the Higgs sector of the SM at the mass ($\sqrt{2}m$) of the field being integrated out, $H$. Since we do not observe the right handed scalars, or a set of SU$(2)_R$ gauge bosons, we must have a large hierarchy of scales between the left and right handed sectors of the model, the simplest way to achieve this is to simply set $\sigma \ll \lambda$.
After integrating out $H$ we find
\begin{equation}
\label{low E tree level}
V_{eff}^{(t)}(h) = \frac{1}{2} \frac{\sigma m^2}{\lambda} h^2 - \frac{\sigma}{4}\left(1+\frac{\sigma}{2\lambda}\right)h^4 + \mathcal{O}(h^5).
\end{equation}
Note that, in the limit of small $\sigma$, i.e. $\sigma \rightarrow 0$ the scalar potential \eqref{scalar pot} has a global $\text{O}(8)$ symmetry which is spontaneously broken down to $\text{O}(7)$, resulting in 7 would-be Goldstone bosons. Three of them become the longitudinal degrees of freedom of the right gauge bosons associated with the $\text{SU}(2)_R \times \text{U}(1)_{B-L}/\text{U}(1)_Y$ coset space. The remaining four form the electroweak Higgs doublet. The pseudo-Goldstone nature of the electroweak Higgs field, $h$, is reflected by the fact that the tree level potential \eqref{low E tree level} vanishes as $\sigma \rightarrow 0$. Hence, a relatively small $\sigma$ is realised naturally.

At $\sqrt{2}m$ we would like to match these couplings with those of the scalar sector defined below the mass threshold (in unitary gauge)
\begin{equation}
V_{eff}^{(t)}(h) = \frac{m_{eff}}{2}^2 h^2 + \frac{\lambda_{eff}}{8}h^4,
\end{equation}
leading to $\lambda_{eff} \approx - 2 \sigma$. Now we seem to have a problem. Recall that to achieve the asymmetric vacuum state we need $\sigma > 0$, but we require $\lambda_{eff} > 0$ for a stable Higgs vacuum so, at tree-level, the low energy limit of the scalar sector of this left-right symmetric model cannot be consistently matched with the Higgs sector of the SM. However, since we are working in the limit $\sigma \ll \lambda$ it is possible that radiative corrections to the potential \eqref{physical potential}, arising from interactions between $h$ and the heavier fields, may dominate in the final expression for the potential. In the next section we will calculate the one loop effective potential for the theory for two reasons - firstly, to see if our model can be matched to the SM at one-loop and secondly, to see if it is possible to generate a non-zero VEV for the $SU(2)_L$ scalars via radiative corrections.

\subsubsection{Yukawa couplings}

Integration over scalar fields does not affect the Yukawa couplings, so we do not need to worry about matching these at the mass threshold of the heavy right handed scalar. At each heavy fermion threshold the story is different - below the scale $M_i$ we integrate out the i'th vector-like fermion, $F_{iJ}$ (J = L,R) resulting in a matching condition for the Yukawa couplings. Under the assumption of a vanishing Cabbibo angle we have 
\begin{equation}
\label{full Yukawa couplings}
\frac{y_{F_i}h}{\sqrt{2}}\left( \overline{f_{iL}} F_{iR} + \overline{F_{iR}} f_{iL}  \right) + y_{F_i}\left(v_R + \frac{H}{\sqrt{2}}\right) \left( \overline{f_{iR}} F_{iL} + \overline{F_{iL}} f_{iR} \right) + M_{F_i}\left( \overline{F_{iL}}F_{iR} + \overline{F_{iR}}F_{iL} \right).
\end{equation}
Thus, we match the Yukawa couplings at $\mu = M_{F_i}$ as
\begin{equation}
\label{Yukawa matching condition}
y_{f_i} = y_{F_i}^2 \frac{ v_R}{M_{F_i}}.
\end{equation}

\subsection{One Loop Effective Potential}

It turns out that one loop matching of our model with the SM electroweak theory is only possible with heavy fermions in the theory, but we have exactly this with the vector-like \eqref{vector-like}. However, there are two competing conditions to worry about, those of matching and vacuum stability - below we will see that consistent matching of our model with the GWS theory can only occur if the Yukawa couplings in the full theory are large, but not too large, otherwise this destabilises the vacuum. We neglect the contribution to the effective potential coming from the right handed gauge bosons since we find it, numerically, to be much smaller than the fermionic contribution at the scales concerned. When we come to the calculation of the threshold corrections we will match the one-loop one \emph{light} particle irreducible (1LPI) effective action for the light field in the full theory:  $\gamma^{(1-loop)}[h]$, with the one-loop one particle irreducible (1PI) effective action for the light field in a low energy effective theory in which the heavy field does not appear: $\Gamma^{(1-loop)}_W [h]$. I.e. we match by choosing our renormalisation conditions for the low energy theory such that
\begin{equation}
\gamma^{(1-loop)}[h] = \Gamma^{(1-loop)}_W[h].
\end{equation}
The threshold conditions appear as finite terms in the renormalisation of the tree-level couplings which are not accounted for in the matching of the one-loop parts of the effective actions for the low and high energy theories.

$\gamma^{(1-loop)}[h]$  is obtained from the full one-loop 1PI effective action, $\Gamma^{(1-loop)}[h,H]$ by integrating the heavy field out according to it's tree-level equations of motion \cite{Burgess:2007pt}
\begin{equation}
\gamma^{(1-loop)}[h] = \Gamma^{(1-loop)}[h,H_t].
\end{equation}
We want the threshold correction to the scalar quartic coupling, arising from integrating out the heavy right handed scalar at the scale $\sqrt{2}m$. Since we include fermions, we will find that, at the one loop level, there is wavefunction renormalisation for the scalar fields. However in the full theory, the one loop corrections to the \emph{light} scalar propagator, with fermions in the loop, are the same as those found for the low energy theory (only at two loops do we find some difference between the sets of diagrams) since the fermionic content for the two theories being matched (at the scalar threshold) is the same. Thus, at the one loop level,  we can match non-kinetic terms in the effective action, i.e., we compare the effective potentials to find the threshold correction to the quartic coupling.
\begin{equation}
V^{(1-loop)}_{W}(h) = V^{(1-loop)}_{eff}(h,H_t).
\end{equation}

Following \cite{Jackiw:1974cv} the one loop contribution to the scalar effective potential, coming from scalars and fermions, obtained from dimensional regularisation and renormalised in the $\overline{\text{MS}}$ scheme is
\begin{equation}
\label{general effective potential}
V^{(1)}_{eff}(h,H) = \frac{1}{64 \pi^2} \sum_i D_i M_i^4 \left( \ln{\frac{M_i^2}{\mu^2}} -\frac{3}{2} \right),
\end{equation}
where $M_i(h,H)$ is the field dependent mass of the i'th particle (in the basis in which the mass matrix is diagonal), coupled to $H$ and/or $h$, with these fields at their respective VEVs. $D_i$ is the number of degrees of freedom for the i'th particle. After integration over $H$, the removal of all non-renormalisable terms and all terms of higher than linear order in $\sigma$, we find the scalar mass (squared) eigenvalues
\begin{gather}
\label{big scalar mass}
M^2_S = 2m^2 - \frac{\sigma}{2}h^2 +\frac{9\lambda^2}{16m^2}h^4,\\
\label{small scalar mass} 
m^2_S =\frac{\sigma}{\lambda}m^2 -\frac{3\sigma}{2}h^2 + \frac{3\lambda^2}{16m^2}h^4.
\end{gather}

As mentioned at the beginning of this section, we need to be careful with the size of the Yukawa couplings in the theory. If we set the vector-like fermion masses too low, the Yukawa couplings are suppressed by the matching condition \eqref{Yukawa matching condition} and turn out to be too small to match the quartic coupling consistently. On the other hand, if we set the mass of the lightest vector-like fermion, $M_T$ to be greater than $v_R$ then we find that the Yukawa couplings are too large and the electroweak vacuum is unstable. Numerically, we find the only solution is to set $M_T < v_R \approx \sqrt{2}m < M_B < M_\tau < \ldots < M_N $. Since the hierarchy of masses is set by \eqref{hierarchy} $M_T$ cannot be too small, otherwise some of the other vector-like fermions would be less massive than $v_R$. Then the only significant fermionic contribution to the threshold correction at $\sqrt{2}m$ comes from the Yukawa coupling, $y_t$, between the top quark and it's partner vector-like fermion, $T$, since this is the the only Yukawa coupling of order one at $m \approx v_R$. In this case the fermion mass eigenvalues are
\begin{gather}
\label{big Yukawa mass}
M_F^2 = M_T^2 + \frac{m^2 y_T^2 }{ \lambda} - \frac{y_T^2}{2}\left( 1 + \frac{\sigma}{\lambda} \right)h^2 + \frac{(7\lambda + 6\sigma)y_t^2}{16 m^2}h^4,\\
\label{small Yukawa mass} 
m_F^2= \frac{y_T^2}{2}h^2 - \frac{y_T^2 \lambda}{4m^2}h^4.
\end{gather}
With \eqref{big scalar mass}, \eqref{small scalar mass}, \eqref{big Yukawa mass} and \eqref{small Yukawa mass} in \eqref{general effective potential} we obtain the one-loop effective potential
\begin{dmath}
\label{full effective potential}
V_{eff}^{(1-loop)}(h) = \frac{1}{2} \frac{\sigma m^2}{\lambda} h^2 - \frac{\sigma}{4}\left(1+\frac{\sigma}{2\lambda}\right)h^4 + \frac{1}{64\pi^2}\left\{ 4m^4\left( \ln{\left[\frac{2m^2}{\mu^2}\right]}-\frac{3}{2} \right) +2\sigma m^2 h^2\left( 1 - \ln{\left[\frac{2m^2}{\mu^2}\right]} \right) +\frac{9\lambda^2}{4}h^4\left( \ln{\left[\frac{2m^2}{\mu^2}\right]}-1 \right) + \left( \frac{\sigma}{\lambda}m^2 - \frac{3\sigma}{2}h^2 \right)^2 \left( \ln{\left[  \frac{ \frac{\sigma}{\lambda}m^2 - \frac{3\sigma}{2}h^2}{\mu^2} \right]} \right) -12\left[ \left(M_T^2 + \frac{m^2 y_T^2 }{ \lambda}\right)^2 \left( \ln{\left[ \frac{M_T^2 + \frac{m^2 y_T^2 }{ \lambda}}{\mu^2} \right]} -\frac{3}{2} \right) + \left(M_T^2 + \frac{m^2 y_T^2 }{ \lambda}\right)y_T^2 h^2 \left( 1-  \ln{\left[ \frac{M_T^2 + \frac{m^2 y_T^2 }{ \lambda}}{\mu^2} \right]} \right) + y_T^4 h^4 \left( \frac{9}{8} \ln{\left[ \frac{M_T^2 + \frac{m^2 y_T^2 }{ \lambda}}{\mu^2} \right]} -\frac{3}{4} \right) + \left(\frac{y_T^2 }{2}h^2 -\frac{y_T^2 \lambda}{4m^2}h^4 \right)^2 \left( \ln{\left[ \frac{\frac{y_T^2 }{2}h^2 -\frac{y_T^2 \lambda}{4m^2}h^4}{\mu^2}\right]} -\frac{3}{2} \right)  \right] \right\}
\end{dmath}

\subsection{One Loop Matching Condition}

We now match terms in the effective potential \eqref{full effective potential} with those from the effective potential of a low energy effective theory, valid between $M_T < \mu < \sqrt{2}m$, and which follows from the Lagrangian with tree-level potential (in unitary gauge again)
\begin{equation}
V_W(h)=\frac{m_{eff}}{2}^2h^2 +\frac{\lambda_{eff}}{8}h^4 + \frac{(y_T)_{eff}h}{\sqrt{2}}\left( \overline{t_L}T_R + \overline{T_R}t_L \right) + \frac{m_R}{\sqrt{2}}\left( \overline{T_L}t_R + \overline{t_R}T_L \right) + M_T\left( \overline{T_L}T_R + \overline{T_R}T_L \right)
\end{equation}
We find the mass (squared) eigenvalues from the Yukawa mass matrix
\begin{gather}
\label{big effective Yukawa mass}
(M_F)_W^2 = M_T^2 + \frac{m^2 y_T^2 }{ \lambda} - \frac{y_T^2}{2}\left( 1 + \frac{\sigma}{\lambda} \right)h^2 + \frac{(7\lambda + 6\sigma)y_T^2}{16 m^2}h^4,\\
\label{small effective Yukawa mass} 
(m_F)_W^2= \frac{(y_T)_{eff}^2}{2}h^2 - \frac{(y_T)_{eff}^2}{4m_R^2}h^4.
\end{gather}
Comparing \eqref{small Yukawa mass} with \eqref{small effective Yukawa mass} gives$(y_T)_{eff} = y_T$ and $m_R = \frac{2m^2y_T^2}{\lambda}$.
Thus, the one loop effective potential for the effective theory is
\begin{dmath}
\label{Wilsonian effective potential}
(V_W)_{eff}^{1-loop}(h)=\frac{m_{eff}}{2}^2h^2 +\frac{\lambda_{eff}}{8}h^4 + \frac{1}{64\pi^2} \left\{ \left( m_{eff}^2 + \frac{3\lambda_{eff}}{2}h^2 \right)^2 \left( \ln{\left[ \frac{m_{eff}^2 + \frac{3\lambda_{eff}}{2}h^2}{\mu^2} \right]} -\frac{3}{2} \right) - 12\left[   \left(M_T^2 + \frac{m^2 y_T^2 }{ \lambda}\right)^2 \left( \ln{\left[ \frac{M_T^2 + \frac{m^2 y_T^2 }{ \lambda}}{\mu^2} \right]} -\frac{3}{2} \right) + y^4 h^4 \left( \frac{1}{2} \ln{\left[ \frac{M_T^2 + \frac{m^2 y_T^2 }{ \lambda}}{\mu^2} \right]} -\frac{1}{2} \right) + \left(\frac{y_T^2 }{2}h^2 -\frac{y_T^2 \lambda}{4m^2}h^4 \right)^2 \left( \ln{\left[ \frac{\frac{y_T^2 }{2}h^2 -\frac{y_T^2 \lambda}{4m^2}h^4}{\mu^2}\right]} -\frac{3}{2} \right)  \right] \right\}
\end{dmath}
Matching \eqref{full effective potential} with \eqref{Wilsonian effective potential} we obtain the threshold correction to the quartic coupling
\begin{dmath}
\label{quartic coupling matching}
\frac{\lambda_{eff}}{8} = -\frac{\sigma}{4} - \frac{9 \lambda^2}{256\pi^2}\left( 1- \ln{\left[ \frac{2m^2}{\mu^2}  \right]}  \right) + \frac{3y^4}{16\pi^2}\left( \frac{1}{4} - \frac{5}{8}  \ln{\left[ \frac{M_T^2 + \frac{m^2 y_T^2 }{ \lambda}}{\mu^2} \right]} \right),
\end{dmath}
with the effective coupling $\lambda_{eff}$ required to be positive.

The quartic coupling also receives a threshold correction each time we integrate out a heavy fermion, however, these corrections are always suppressed by a loop factor and, as mentioned above, the tree level matching at a fermion threshold is $\lambda_{eff} = \lambda$. Thus, we ignore loop threshold corrections to $\lambda$ at heavy fermion thresholds. 

To summarise, when we solve the RGEs we will match the couplings as follows: at each heavy fermion threshold, $M_{F_i}$, the only coupling to receive a threshold correction is $y_{f_i}$, according to \eqref{Yukawa matching condition}. At the right handed scalar threshold, $\sqrt{2}m$, the only coupling to receive a threshold correction is $\lambda$, according to \eqref{quartic coupling matching}. Finally, at $v_R$ we match the gauge couplings using \eqref{U(1) matching}

\subsection{Minimisation of the Potential}

Before solving the RGEs, we need to check that the effective potential \eqref{full effective potential} now has a minimum for $h \neq 0$. Note that, to one loop order, the potential \eqref{full effective potential} (obtained by solving $\frac{\partial V_t}{\partial H} = 0$ at tree level) agrees with the one obtained by calculating the full effective potential for $h$ and $H$, and then solving $\frac{\partial V^{1-loop}_{eff}(h,H)}{\partial H} = 0$ \cite{Burgess:2007pt}. In our small $\sigma$ limit the potential depends mainly on $y_T$ and $\lambda$, so we now drop all terms in $\sigma$ in the effective potential. Also, we will find $M_T \ll m^2 y^2_T/\lambda$ enabling us to remove $M_T$. In addition, we expect radiative corrections to give only a small correction to $\langle h \rangle \equiv h$, and so we also drop factors of $h/m$ to find the transcendental equation
\begin{equation}
\label{electroweak minimum}
h^2 = \frac{2 (m^2/\lambda) \left(  \ln{\left[ \frac{ m^2y_T^2}{\lambda \mu^2} \right]} -1\right)}{\frac{3\lambda^2}{4y_T^4}\left( 1 - \ln{\left[ \frac{2m^2}{\mu^2} \right]} \right) + \left(\frac{9}{2} \ln{\left[ \frac{ m^2y_T^2}{\lambda \mu^2} \right]} -3 \right) + \left( \ln{\left[ \frac{y_T^2 h^2}{2\mu^2} \right]} -1 \right) } .
\end{equation}
Setting
\begin{equation}
\mu^2 = \frac{m^2}{\lambda y_T ^2}\left( 1 - \frac{h^2}{v_R^2} \right)^{-1}e^{-1}
\end{equation}
we find 
\begin{dmath}
h^2 \approx \frac{2 (m^2/\lambda) \ln{\left[ 1-\frac{h^2}{v_R^2}\right]}}{\ln{\left[ \frac{y_T^4 \lambda h^2}{2m^2}  \right]}+\frac{3}{2}  -\frac{3\lambda^2}{4y_T^4} \ln{\left[ 2 \lambda y_T^2 \right]} } .
\end{dmath}
For this choice of $\mu$ we find that the extremum in $V(h)$ is a minimum, and numerically we find that $h \equiv v_{EW}$ is many orders of magnitude smaller than $v_R$, as expected.

\subsection{Solution to the RGEs}

\begin{figure}[h!]
\centering
\subfloat[Running gauge couplings.]{\label{fig:LRSymGsMT4_710to9GeVH125MR10to10GeV}\includegraphics[width=0.49\textwidth]{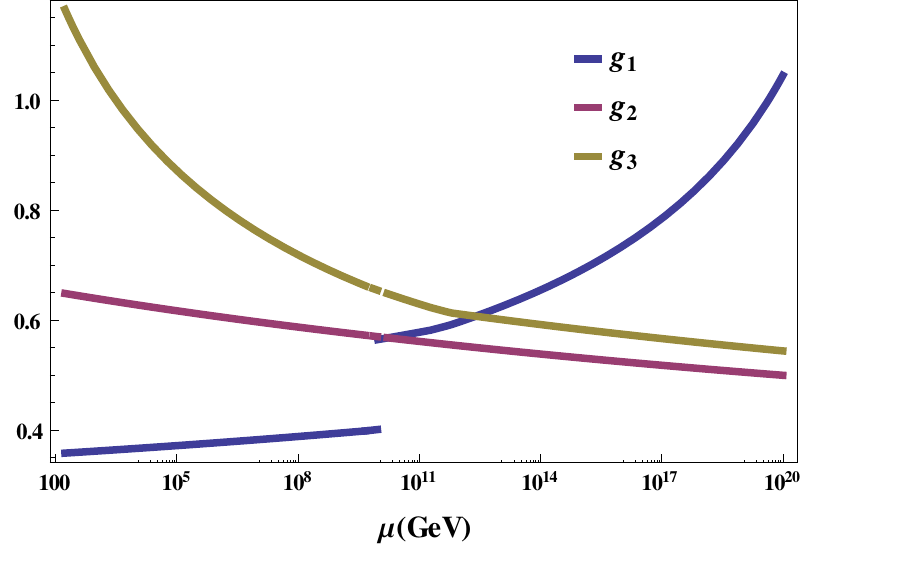}}
\subfloat[Running Yukawa couplings.]{\label{fig:LRSymYsMT4_710to9GeVH125MR10to10GeV}\includegraphics[width=0.49\textwidth]{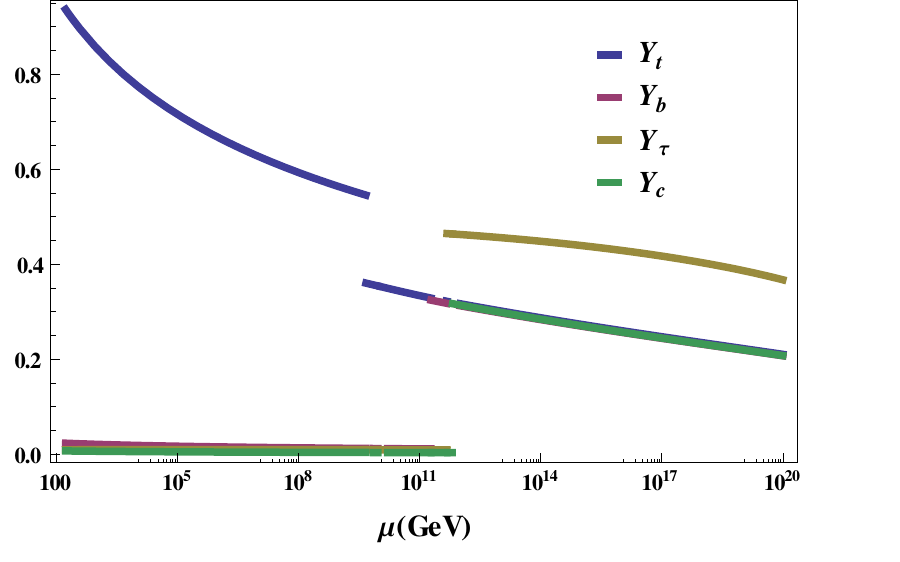}}\qquad
\subfloat[Running Higgs quartic coupling]{\label{fig:LRSymLambdaMT4_710to9GeVH125MR10to10GeV}\includegraphics[width=0.49\textwidth]{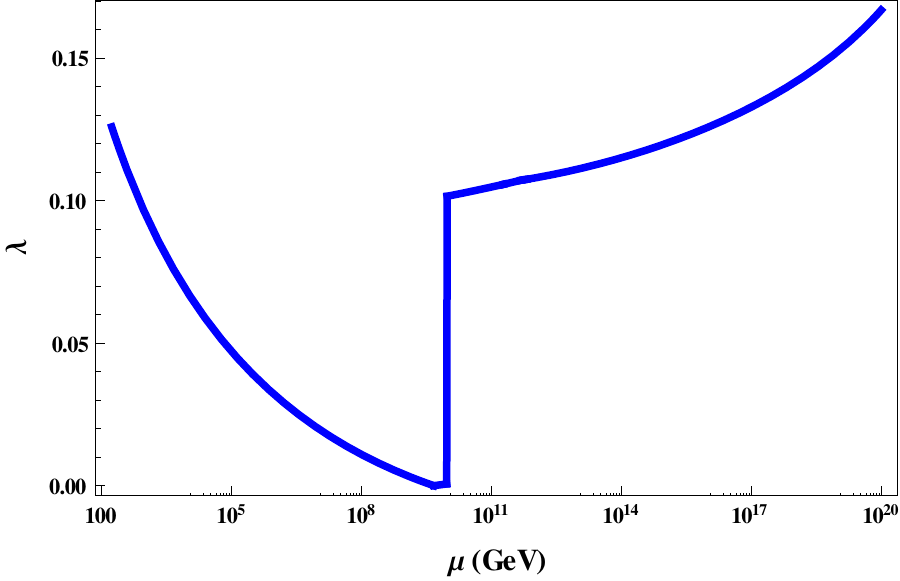}}
\caption[]{Running couplings in the ALRSM with $M_T = 4.7 \times 10^9$ GeV and $m = 1 \times 10^{10}$ GeV.}
\label{fig:LRSym}
\end{figure}

Since we're working with the RGEs obtained from dimensional regularisation in the $\overline{\text{MS}}$ scheme, we require to a set of $\beta$-functions for each effective field theory in the range of the running (from the electroweak scale up to $M_P$). Here, the SM RGEs are solved to two loops and all other RGEs to one loop order. Unusually for this type of analysis, we have chosen to run the Yukawa couplings for the b and c quarks and the $\tau$ because the relations \eqref{Yukawa matching condition} and \eqref{hierarchy} make the Yukawa coupling $y_{F_i}$ of order $y_T$ at the particle threshold $M_{F_i}$, so we must include their contribution to the running of $\lambda$. Since the vector-like fermion masses follow a hierarchy inverse to that of their SM partners, the lightest, in order of increasing mass, are $T, B, \boldsymbol{\tau}, C$ and we include the Yukawa couplings for these in the running. The S does not appear until a couple of orders of magnitude later than the $\boldsymbol{\tau}$ and $C$, and too close to the Planck scale to affect the running of $\lambda$ greatly, so we ignore the Yukawa coupling for this and any heavier fermions. As a (happy) consequence of staggering the fermion masses in this way, we also find that the $\beta$-function for $g_{B-L}$ is not increased too greatly by the additional $(B-L)$ charge of all the extra fermionic states, and we avoid a Landau pole that we would have otherwise run into if all the extra fermions had $M_{F_i} \approx v_R$. The $\beta$-functions for each effective theory, starting with the SM and finishing with the full theory, appear in appendix \ref{app:LR beta functions}.

Boundary conditions (at the electroweak scale) for the couplings $g_1$, $g_2$, $g_3$, $\lambda_h$ and $y_t$ can be found in Appendix \ref{app:Initial Conditions}, whilst those for $y_b$, $y_\tau$ and $y_c$ are approximated by
\begin{equation}
y_{f_i} = \frac{m_{f_i}}{m_t}y_t.
\end{equation}
We match all couplings in the $\theta$-approximation. $\lambda$ is matched with $\lambda_{eff}$ at $\mu = \sqrt{2}m$, in which case one sees from \eqref{quartic coupling matching} that, in order to obtain a real and positive $\lambda$, we must have $\lambda_{eff}$ very small and $y_t$ a reasonable size. As a consequence, we find that matching is only possible if we set the heavy right handed scalar threshold just below the instability scale and $M_T$ an order of magnitude further below at most. Also, as mentioned above, if $y_T$ is too large the vacuum becomes unstable, so we find a relatively small window of opportunity for $M_T$ and $\sqrt{2}m$: $\frac{\sqrt{2}m}{10}  < M_T < \sqrt{2}m \lesssim \mu_I$. Taking into account the above, we find it possible to rescue stability of the electroweak vacuum in the ALRSM. 

Solutions to the RGEs for a particular set of parameters are depicted in Figure \ref{fig:LRSym}. We can account for various different positive contributions to $\beta_\lambda$ in this example. Firstly, a threshold correction arising from the matching of the $\text{U}(1)_Y$ and $\text{U}(1)_{B-L}$ gauge couplings at the scale of left-right symmetry breaking, $v_R$ (see \eqref{U(1) matching}), increases the value of the Abelian gauge coupling (see Figure \ref{fig:LRSymGsMT4_710to9GeVH125MR10to10GeV}), and hence gives an increased positive contribution to $\beta_\lambda$ at high energies, $\mu > v_R$. At the same time the threshold correction to $\lambda$ at $\mu = \sqrt{2}\pi \approx v_R$ from integrating out the right scalar doublet $\phi_R$ (see \eqref{quartic coupling matching}) significantly increases $\lambda$. Finally, additional contributions to $\beta_\lambda$ from $\phi_L-\phi_R$ interactions above $v_R$ render $\lambda$ increasing at high energies. Hence, the development of vacuum instability is avoided as can be seen in Figure \ref{fig:LRSymLambdaMT4_710to9GeVH125MR10to10GeV}. Note, however, that even though matching of the top Yukawa coupling with $y_T$ decreases the coupling above the matching scale $\mu = M_T$ since $M_T < v_R$ (see \eqref{Yukawa matching condition}), other Yukawa couplings are increased by the same process, since we assume a 'natural' seesaw hierarchy for the heavy vector-like fermions (see Figure \ref{fig:LRSymYsMT4_710to9GeVH125MR10to10GeV}) so the overall negative contribution to $\beta_\lambda$ from the Yukawa couplings is roughly the same as in the SM.  Whilst our concrete numerical example is given just to illustrate the different possible ways to solve the instability problem, it is clear that there exits a range of parameters for which absolute vacuum stability can be maintained. A detailed analysis of the allowed parameter space is beyond the scope of this paper.

\section{Conclusion}

The observed mass of the Higgs particle, $m_h=125-126$ GeV, implies that, within the SM, the electroweak vacuum is metastable, and within the inflationary scenario it may even be unstable.  This may be a hint for new physics beyond the SM, which maintains absolute stability of the vacuum state. In this paper we have discussed which modifications in the gauge, Higgs and Yukawa sectors at high energies can fulfil this role. In particular, we have concentrated on new physics related to the mechanism of neutrino mass generation. The Higgs vacuum stability problem cannot be solved within the simplest type-I seesaw model where only right-handed neutrinos are added to the SM. In the type-III seesaw model the vacuum can be stabilised but only for a light triplet of fermions, which contribute to the running of the Higgs quartic coupling indirectly, through the running of $SU(2)_L$ gauge coupling.

Then we have considered in detail two extensions of the SM Ð the type-II seesaw model and the left-right symmetric model with the universal seesaw mechanism for all quarks and leptons. In the type-II seesaw model we have solved the two-loop RGEs for electroweak Higgs quartic coupling and other relevant couplings with tree-level threshold corrections properly incorporated. Threshold corrections turn out to be important for large triplet scalar masses $m_\Delta$. We found that for a wide range of parameters one can obtain an absolutely stable vacuum (see, Figure 2).

In the left-right symmetric model, one typically employs a bi-doublet of scalar fields alongside the left and right doublets of scalars to generate masses for all the SM fermions. In the ALRSM, we have considered in this paper, only left and right doublets are introduced whilst the masses of ordinary quarks and leptons are generated by the universal seesaw mechanism via the tree-level exchange of additional heavy vector-like fermions. First, we have shown that the ALRSM is a consistent theory on its own, and nor extra scalars or an explicit violation of left-right symmetry is required, as was assumed in previous works. Interestingly, the ordinary electroweak symmetry of the SM is broken when quantum corrections are taken into account. We have demonstrated that the SM and ALRSM can be consistently matched at high-energy scales by computing explicitly the one-loop matching conditions. As a result, we found that modification of the $\beta_{\lambda}$-function for the Higgs quartic coupling stems from modifications of the hypercharge gauge coupling, Yukawa couplings and scalar interaction couplings. These collectively drive $\beta_{\lambda}$ to be positive, and hence the Higgs quartic coupling to increase at high energies, making the Higgs vacuum absolutely stable (see, Figure 4).  
        
\paragraph{Acknowledgments.} We thank Kristian McDonald for useful discussions. This work was partially supported by the Australian Research Council.

\appendix

\section{$\overline{\text{MS}}$ Standard Model Couplings at the Electroweak Scale}
\label{app:Initial Conditions}

In order to solve the RGEs we require boundary conditions for the SM couplings at the electroweak scale. All values used were taken from \cite{Degrassi:2012ry} and are as follows. The gauge couplings at the $Z$ pole-mass $m_Z = 91.1875$ GeV \cite{ALEPH:2005ab} are
\begin{gather}
\frac{g_1^2(m_Z)}{4\pi} = \alpha_Y^{-1}(m_Z)=98.350,\\
\frac{g_2^2(m_Z)}{4\pi} = \alpha_2^{-1}(m_Z)=29.587,\\
\frac{4\pi}{g_3^2(m_Z)} = \alpha_3(m_Z)=0.1184.
\end{gather}
The $\overline{\text{MS}}$ top Yukawa and Higgs quartic couplings are matched to the top and Higgs pole-masses according to
\begin{dmath}
y_t(m_t) = 0.93587 + 0.00557\left(\frac{m_t}{\text{GeV}} - 173.15 \right) - 0.00003 \left(\frac{m_h}{\text{GeV}} - 125 \right) - 0.00041 \left(\frac{\alpha_3(m_Z)-0.1184}{0.0007} \right)
\end{dmath}
\begin{dmath}
\lambda_h(m_t) = 0.12577 + 0.00205 \left(\frac{m_h}{\text{GeV}} -125 \right) - 0.00004 \left(\frac{m_t}{\text{GeV}} - 173.15 \right)
\end{dmath}

\section{Standard Model Beta Functions}
\label{app:SM beta functions}

The following were taken from \cite{Holthausen:2011aa}. We expand the $\beta$-function for a generic coupling, $\lambda_i$, to ($n$th) loop order as
\begin{equation}
\beta_{\lambda_i} = \sum_n \frac{1}{(64\pi^2)^n}\beta_{\lambda_i}^{(n)},
\end{equation}
then
\begin{gather}
\beta _{g_1}^{(1)} = \frac{41}{6}g_1^3, \\
\beta _{g_1}^{(2)} = g_1^3\left(\frac{199}{18}g_1^2+\frac{9}{2}g_2^2+\frac{44}{3}g_3^2-\frac{17}{6}y_t^2\right), \\
\beta _{g_2}^{(1)}=-\frac{19}{6}g_2^3, \\
\beta _{g_2}^{(2)}=g_2^3\left(\frac{3}{2}g_1^2+\frac{35}{6}g_2^2+12g_3^2-\frac{3}{2}y_t^2\right),\\
\beta _{g_3}^{(1)}=-7g_3^3, \\
\beta _{g_3}^{(2)}=g_3^3\left(\frac{11}{6}g_1^2+\frac{9}{2}g_2^2-26g_3^2-2y_t^2\right), \\
\beta _{y_t}^{(1)}=y_t\left(\frac{9}{2}y_t^2+\frac{3}{2}y_b^2+\frac{3}{2}y_c^2+y_{\tau }^2\right)+y_t\left(-\frac{17}{12}g_1^2-\frac{9}{4}g_2^2 -8g_3^2\right), \\
\beta _{y_c}^{(1)}=y_c\left(\frac{9}{2}y_c^2+\frac{3}{2}y_b^2+\frac{3}{2}y_t^2+y_{\tau }^2\right)+y_c\left(-\frac{17}{12}g_1^2-\frac{9}{4}g_2^2 -8g_3^2\right), \\
\beta _{y_b}^{(1)}= y_b\left(\frac{9}{2}y_b^2+\frac{3}{2}y_t^2+y_{\tau }^2\right)+y_b\left(-\frac{5}{12}g_1^2-\frac{9}{4}g_2^2 -8g_3^2\right), \\
\beta _{y_{\tau }}^{(1)}=y_{\tau }\left(\frac{5}{2}y_{\tau }^2+3y_t^2+3y_b^2\right)+y_{\tau }\left(-\frac{15}{4}g_1^2-\frac{9}{4}g_2^2 \right),
\end{gather}
\begin{dmath}
\beta _{y_t}^{(2)}=y_t\left(-12y_t^4+y_t^2\left(\frac{131}{16}g_1^2+\frac{225}{16}g_2^2+36g_3^2-12\lambda_h \right) \\ +\frac{1187}{216}g_1^4-\frac{3}{4}g_1^2g_2^2+\frac{19}{9}g_1^2g_3^2-\frac{23}{4}g_2^4+9g_2^2g_3^2-108g_3^4+6\lambda_h ^2\right), 
\end{dmath}
\begin{dmath}
\beta _{\lambda_h }^{(1)}=\lambda_h  \left(-9g_2^2-3g_1^2+12\left(y_t^2+y_b^2+\frac{y_{\tau }^2}{3}+y_c^2\right)\right)+24\lambda_h ^2+\frac{3}{4}g_2^4+\frac{3}{8}\left(g_1^2+g_2^2\right)^2-6\left(y_t^4+y_b^4+y_c^4+\frac{y_{\tau }^4}{3}\right),
\end{dmath}
\begin{dmath}
\beta _{\lambda_h }^{(2)}=-312\lambda_h ^3-144y_t^2\lambda_h ^2+36\lambda_h ^2\left(g_1^2+3g_2^2\right)-3\lambda_h y_t^4+\lambda_h y_t^2\left(\frac{85}{6}g_1^2+\frac{45}{2}g_2^2+80g_3^2\right)-\frac{73}{8}\lambda_h g_2^4+\frac{39}{4}\lambda_h g_2^2g_1^2+\frac{629}{24}\lambda_h g_1^4+30y_t^6-32g_3^2y_t^4-\frac{8}{3}y_t^4g_1^2-\frac{9}{4}y_t^2g_2^4+\frac{21}{2}y_t^2g_2^2g_1^2-\frac{19}{4}y_t^2g_1^4+\frac{305}{16}g_2^6-\frac{289}{48}g_2^4g_1^2-\frac{559}{48}g_2^2g_1^4-\frac{379}{48}g_1^6.
\end{dmath}

\section{Type-II Seesaw Beta Functions ($\mu > m_\Delta$)}
\label{app:TII beta functions}

The following were taken from \cite{Schmidt:2007nq}
\begin{gather}
\beta_{g_1}^{(1)} = \frac{47}{6}g_1^3,\\
\beta_{g_2}^{(1)}=-\frac{5}{2}g_2^3,\\ 
\beta_{\lambda }^{(1)}=24\lambda ^2-3\lambda \left(3g_2^2+g_1^2\right)+\frac{3}{4}g_2^4+\frac{3}{8}\left(g_1^2+g_2^2\right)^2+12\lambda y_t^2-6y_t^4+3\lambda _4^2+2\lambda _5^2,\\ 
\beta _{y_\Delta }^{(1)}=3y_{\Delta }^3+y_{\Delta }\left(-\frac{3}{2}\left(g_1^2+3g_2^2\right)+y_{\Delta }^2\right),
\end{gather}
\begin{multline}
\beta _{\lambda _1}^{(1)}=-12g_1^2\lambda _1-24g_2^2\lambda _1+12g_1^4+18g_2^4+24g_1^2g_2^2\\ +14\lambda _1^2+4\lambda _1\lambda _2+2\lambda _2^2+4\lambda _4^2+4\lambda _5^2+4 y_{\Delta }^2\lambda _1-8y_{\Delta }^4, 
\end{multline}
\begin{gather}
\beta _{\lambda _2}^{(1)}=-12g_1^2\lambda _2-24g_2^2\lambda _2+12g_2^4-48g_1^2g_2^2+3\lambda _2^2+12\lambda _1\lambda _2+8\lambda _5^2+4 y_{\Delta }^2\lambda _2+8y_{\Delta }^4,\\ 
\beta _{\lambda _4}^{(1)}=-\frac{15}{2}g_1^2\lambda _4-\frac{33}{2}g_2^2\lambda _4+3g_1^4+6g_2^4+\left(8\lambda _1+2\lambda _2+12\lambda +4\lambda _4+6y_t^2+2y_{\Delta }^2\right)\lambda _4+8\lambda _5^2,\\ 
\beta _{\lambda _5}^{(1)}=-\frac{15}{2}g_1^2\lambda _5-\frac{33}{2}g_2^2\lambda _5-6g_1^2g_2^2+\left(2\lambda _1-2\lambda _2+4\lambda +8\lambda _4+6y_t^2+2y_{\Delta }^2\right)\lambda _5,\\ 
\beta _{\lambda _6}^{(1)}= \lambda _6 \left(4\lambda -4\lambda _4+8\lambda _5-\frac{9}{2}g_1^2-\frac{21}{2}g_2^2+6y_t^2+y_{\Delta }^2\right).
\end{gather}

\section{Left-Right Symmetric Model Beta Functions}
\label{app:LR beta functions}

All $\beta$-functions for this section were obtained using the generic formulae available in \cite{Machacek:1983tz, Machacek:1983fi, Machacek:1984zw}.

\subsection{Beta functions for $M_T < \mu < \sqrt{2}m$}

Let $\lambda_i = \{g_1,g_2,g_3,y_T,y_b,y_c, y_\tau\}$, then
\begin{gather}
\beta_{\lambda_i}^{(1)} =  \sum_{j=1}^7 a_{ij} \lambda_j^2 \lambda_i, \ \ \
a_{ij} = \left(\begin{array}{ccccccc}\frac{155}{18} & 0 & 0 & 0 & 0 & 0 & 0 \\0 &- \frac{19}{6} & 0 & 0 & 0 & 0 & 0 \\0 & 0 &- \frac{19}{3} & 0 & 0 & 0 & 0 \\-\frac{17}{12} & -\frac{9}{4} & -8 & \frac{9}{2} & \frac{3}{2} & \frac{3}{2} & 1 \\-\frac{5}{12} & -\frac{9}{4} & -8 & \frac{3}{2} & \frac{9}{2} & \frac{3}{2} & 1 \\-\frac{17}{12} & -\frac{9}{4} & -8 & \frac{3}{2} & \frac{3}{2} & \frac{9}{2} & 1 \\-\frac{15}{4} & -\frac{9}{4} & 0 & 3 & 3 & 3 & \frac{5}{2} \end{array}\right),
\end{gather}
\begin{dmath}
\beta_{\lambda_{eff} }^{(1)}=16\lambda_{eff} ^2+3\lambda_{eff} \left(4 \left(y_T^2+y_b^2+\frac{y_{\tau }^2}{3}+y_c^2\right)-3g_2^2-g_1^2\right)+\frac{3}{2}g_2^4+\frac{3}{4}\left(g_1^2+g_2^2\right)^2-12\left(y_T^4+y_b^4+y_c^4+\frac{y_{\tau }^4}{3}\right).
\end{dmath}

\subsection{Beta functions for $\sqrt{2}m < \mu < M_B$}

Let $\lambda_i = \{g_{B-L},g_L,g_3,y_T,y_b,y_c, y_\tau\}$, then
\begin{gather}
\beta^{(1)}_{\lambda_i} =  \sum_{j=1}^7 a_{ij} \lambda_j^2 \lambda_i, \ \ \
a_{ij} = \left(\begin{array}{ccccccc}\frac{43}{9} & 0 & 0 & 0 & 0 & 0 & 0 \\0 & -\frac{19}{6} & 0 & 0 & 0 & 0 & 0 \\0 & 0 & -\frac{19}{3} & 0 & 0 & 0 & 0 \\-\frac{17}{12} & -\frac{9}{4} & -8 & \frac{9}{2} & \frac{3}{2} & \frac{3}{2} & 1 \\-\frac{5}{12} & -\frac{9}{4} & -8 & \frac{3}{2} & \frac{9}{2} & \frac{3}{2} & 1 \\-\frac{17}{12} & -\frac{9}{4} & -8 & \frac{3}{2} & \frac{3}{2} & \frac{9}{2} & 1 \\-\frac{15}{4} & -\frac{9}{4} & 0 & 3 & 3 & 3 & \frac{5}{2} \end{array}\right),
\end{gather}
\begin{dmath}
\beta_{\lambda }^{(1)}=16\lambda ^2+3\lambda \left(4 \left(y_T^2+y_b^2+\frac{y_{\tau }^2}{3}+y_c^2\right)-3g_L^2-g_{B-L}^2\right)+\frac{3}{2}g_L^4+\frac{3}{4}\left(g_{B-L}^2+g_L^2\right)^2-12\left(y_T^4+y_b^4+y_c^4+\frac{y_{\tau }^4}{3}\right).
\end{dmath}

\subsection{Beta functions for $M_B < \mu < M_\tau$}

Let $\lambda_i = \{g_{B-L},g_L,g_3,y_T,y_B,y_c, y_\tau\}$, then
\begin{gather}
\beta^{(1)}_{\lambda_i} =  \sum_{j=1}^7 a_{ij} \lambda_j^2 \lambda_i, \ \ \
a_{ij} = \left(\begin{array}{ccccccc}\frac{59}{9} & 0 & 0 & 0 & 0 & 0 & 0 \\0 & -\frac{19}{6} & 0 & 0 & 0 & 0 & 0 \\0 & 0 & -\frac{17}{3} & 0 & 0 & 0 & 0 \\-\frac{17}{12} & -\frac{9}{4} & -8 & \frac{9}{2} & \frac{3}{2} & \frac{3}{2} & 1 \\-\frac{5}{12} & -\frac{9}{4} & -8 & \frac{3}{2} & \frac{9}{2} & \frac{3}{2} & 1 \\-\frac{17}{12} & -\frac{9}{4} & -8 & \frac{3}{2} & \frac{3}{2} & \frac{9}{2} & 1 \\-\frac{15}{4} & -\frac{9}{4} & 0 & 3 & 3 & 3 & \frac{5}{2} \end{array}\right),
\end{gather}
\begin{dmath}
\beta_{\lambda }^{(1)}=16\lambda ^2+3\lambda \left(4 \left(y_T^2+y_B^2+\frac{y_{\tau }^2}{3}+y_c^2\right)-3g_L^2-g_{B-L}^2\right)+\frac{3}{2}g_L^4+\frac{3}{4}\left(g_{B-L}^2+g_L^2\right)^2-12\left(y_T^4+y_B^4+y_c^4+\frac{y_{\tau }^4}{3}\right).
\end{dmath}

\subsection{Beta functions for $M_\tau < \mu < M_C$}

Let $\lambda_i = \{g_{B-L},g_L,g_3,y_T,y_B,y_c, y_{\boldsymbol{\tau}}\}$, then
\begin{equation}
\beta^{(1)}_{\lambda_i} =  \sum_{j=1}^7 a_{ij} \lambda_j^2 \lambda_i, \ \ \
a_{ij} = \left(\begin{array}{ccccccc}7 & 0 & 0 & 0 & 0 & 0 & 0 \\0 & -\frac{19}{6} & 0 & 0 & 0 & 0 & 0 \\0 & 0 & -5 & 0 & 0 & 0 & 0 \\-\frac{17}{12} & -\frac{9}{4} & -8 & \frac{9}{2} & \frac{3}{2} & \frac{3}{2} & 1 \\-\frac{5}{12} & -\frac{9}{4} & -8 & \frac{3}{2} & \frac{9}{2} & \frac{3}{2} & 1 \\-\frac{17}{12} & -\frac{9}{4} & -8 & \frac{3}{2} & \frac{3}{2} & \frac{9}{2} & 1 \\-\frac{15}{4} & -\frac{9}{4} & 0 & 3 & 3 & 3 & \frac{5}{2} \end{array}\right),
\end{equation}
\begin{dmath}
\beta_{\lambda }^{(1)}=16\lambda ^2+3\lambda \left(4 \left(y_T^2+y_B^2+\frac{y_{\boldsymbol{\tau}}^2}{3}+y_c^2\right)-3g_L^2-g_{B-L}^2\right)+\frac{3}{2}g_L^4+\frac{3}{4}\left(g_{B-L}^2+g_L^2\right)^2-12\left(y_T^4+y_B^4+y_c^4+\frac{y_{\boldsymbol{\tau}}^4}{3}\right).
\end{dmath}

\subsection{Beta functions for $\mu > M_C $}

Let $\lambda_i = \{g_{B-L},g_L,g_3,y_T,y_B,y_C, y_{\boldsymbol{\tau}}\}$, then
\begin{equation}
\beta^{(1)}_{\lambda_i} =  \sum_{j=1}^7 a_{ij} \lambda_j^2 \lambda_i, \ \ \
a_{ij} = \left(\begin{array}{ccccccc}\frac{41}{3} & 0 & 0 & 0 & 0 & 0 & 0 \\0 & -\frac{19}{6} & 0 & 0 & 0 & 0 & 0 \\0 & 0 & -3 & 0 & 0 & 0 & 0 \\-\frac{17}{12} & -\frac{9}{4} & -8 & \frac{9}{2} & \frac{3}{2} & \frac{3}{2} & 1 \\-\frac{5}{12} & -\frac{9}{4} & -8 & \frac{3}{2} & \frac{9}{2} & \frac{3}{2} & 1 \\-\frac{17}{12} & -\frac{9}{4} & -8 & \frac{3}{2} & \frac{3}{2} & \frac{9}{2} & 1 \\-\frac{15}{4} & -\frac{9}{4} & 0 & 3 & 3 & 3 & \frac{5}{2} \end{array}\right),
\end{equation}
\begin{dmath}
\beta_{\lambda }^{(1)}=16\lambda ^2+3\lambda \left(4 \left(y_T^2+y_B^2+y_C^2+\frac{y_{\boldsymbol{\tau}}^2}{3}\right)-3g_L^2-g_{B-L}^2\right)+\frac{3}{2}g_L^4+\frac{3}{4}\left(g_{B-L}^2+g_L^2\right)^2-12\left(y_T^4+y_B^4+y_C^4+\frac{y_{\boldsymbol{\tau}}^4}{3}\right).
\end{dmath}

\end{document}